\documentclass[twocolumn,showpacs,preprintnumbers,amsmath,amssymb]{revtex4}

\usepackage{graphicx}
\usepackage{dcolumn}
\usepackage{bm}
\usepackage[colorlinks]{hyperref}

\begin{document}

\title{A Review of Recent Studies of
 Geographical Scale-Free Networks}

\author{Yukio Hayashi}
\affiliation{
Japan Advanced Institute of Science and Technology,
Ishikawa, 923-1292, Japan}
\author{Jun Matsukubo}
\affiliation{Kitakyusyu National College of Technology, 
Fukuoka 802-0985, Japan}

\date{\today}

\begin{abstract}
The scale-free (SF) structure that commonly appears 
in many complex networks
is one of the hot topics related to social, biological, and information
sciences. 
The self-organized generation mechanisms are expected to be useful for
efficient communication or robust connectivity in socio-technological
infrastructures.
This paper is the first review of geographical SF network models.
We discuss the essential generation mechanisms to induce 
the structure with power-law behavior 
and the properties of planarity and link length.
The distributed designs of geographical SF networks without 
the crossing and long-range links that cause the interference
and dissipation problems are very important 
for many applications such as in 
the Internet, power-grid, mobile, and sensor systems.
\end{abstract}


\maketitle

\section{Introduction}
As a breakthrough in network science \cite{Buchnan02}, 
it has been found \cite{Albert02} 
that many real systems of social, technological, and biological origins 
have the surprisingly common topological structure called 
{\it small-world} (SW) \cite{Watts98} and {\it scale-free}
 (SF) \cite{Barabasi99}. 
The structure is characterized by the SW properties that  
the average path length over all nodes (vertices) is short as similar to 
that in random graphs,  
and that the clustering coefficient,  
defined by the average ratio of the number of links (edges) 
connecting to its nearest neighbors of a node 
to the number of possible links between 
all these nearest neighbors, 
is large as similar to that in regular graphs. 
Large clustering coefficient means the high frequency of 
``the friend of a friend is also his/her friend.''
As the SF property, the degree distribution follows a power-law, 
$P(k) \sim k^{- \gamma}$, $2 < \gamma < 3$;
the fat-tail distribution
consists of many nodes with low degrees and a few hubs with 
very high degrees.
Moreover, a proposal of the universal mechanisms \cite{Barabasi99}
to generate SF networks 
inspired to elucidate the topological properties. 
One of the advantage is that 
SF networks are optimal in minimizing 
both the effort for communication 
and the cost for maintaining the connections \cite{Cancho03}. 
Intuitively, SF network is positioned between  
star or clique graph for minimizing the path length 
(the number of hops or legs)
and random tree for minimizing the number of links
within the connectivity. 
Another important property is that  
SF networks are robust against random failures but vulnerable  
against the targeted attacks on hubs. 
This vulnerability called  
``Achilles' heel of the Internet'' \cite{Albert00a} 
frightened us.
However the vulnerability is a double-edged sword for 
information delivery and spreading of viruses,
we expect that these properties will be useful for
developing efficient and fault-tolerant networks
with a defense mechanism based on the protection of hubs.  
Since 
the SF structure is  at least selected with self-organized manners 
in social and biological environments,
the evolutional mechanisms may give insight to 
distributed network designs or social managements in 
communication or business.

On the other hand, 
in contrast to abstract graphs, many real networks are embedded in a
metric space. 
It is therefore natural to investigate the possibility of 
embedding SF networks in space.
The related applications are very wide in the 
Internet(routers), power-grids, airlines, 
mobile communication \cite{Hong02}, 
sensor networks \cite{Culler04}, and so on.
However most of the works on SF networks were irrelevant to a   
geographical space. 
In this paper, 
focusing on the SF structure found in many real systems,    
we consider generation rules of 
geographical networks whose nodes are set on a Euclidean  
space and the undirected links between them are  
weighted by the Euclidean distance.
  
The organization of this paper is as follows.  
In section 2, we introduce an example that shows 
the restriction of long-range links in real networks.
Indeed, 
the decay of connection probability for the distance between nodes 
follows exponential or power-law.
In section 3, we review recent studies of geographical SF network   
models, which are categorized in three classes by the generation rules.
We refer to the analytical forms of 
degree distributions that characterize the SF structure.
In section 4, 
we consider the relations among these models. 
In addition, we compare the properties of 
planarity and distance of connections.
Finally, in section 5, the summary and further issues 
are briefly discussed.

\section{Spatial distribution in real-world networks}
The restriction of long-range links has been observed in real networks: 
Internet at both router and autonomous system (AS) levels 
obtained by using NETGEO tool to identify the geographical coordinates
of 228,265 routers \cite{Yook02}.
These data suggest that 
the distribution of link lengths (distance)
is inversely proportional to the lengths,  
invalidating the Waxman's exponentially decay rule \cite{Waxman88}  
which is widely used in traffic simulations. 
Other evidence has been reported for the real data of 
Internet as AS level (7,049 nodes and 13,831 links)
compiled by the University of Oregon's Route Views project, 
road networks of US interstate highway (935 nodes and 1,337 links)
extracted from GIS databases, 
and flight-connections (187 nodes and 825 links)
in a major airline \cite{Gastner04}. 
It has been shown that 
all three networks have a clear bias towards shorter links to 
reduce the costs for construction and maintenance, 
however there exist some differences:
the road network has only very short links on the order of 10km to
100km in the sharply decaying distribution, 
while the Internet and airline networks have much longer ones 
in the bimodal distribution with distinct peaks around 
2000km or less and 4000km.
These differences 
may come from physical constraints in the link cost or 
the necessaries of long distant direct connections.

\begin{figure}[htb] 
  \begin{center}
  \includegraphics[width=70mm]{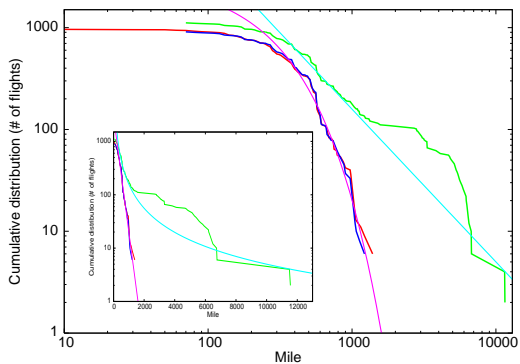}    
  \end{center}
  \caption{Cumulative number of flights in Japanese airlines.
 The red, blue, and green lines correspond to 
 domestic flights in JAL, ANA, and that including international flights
 (Inset: semi-log scale). 
 The magenta and cyan lines guide the estimated functions of 
 exponential and power-law, respectively.} 
  \label{fig_airlines_data}
\end{figure}

As a similar example, 
we investigate the distributions of link lengths 
(distances of flights) in Japanese airlines \cite{JALANA}.
The networks consists of 52 nodes (airports)
and 961 links (flights)
in the Japan AirLines (JAL), 
49 nodes and 909 links in the All Nippon Airlines (ANA), and 
84 nodes and 1,114 links including the international one.
Fig. \ref{fig_airlines_data} shows the cumulative number of flights 
for the decreasing order of length measured by mile.
We remark an exponential decay in the domestic flights 
(red and blue lines in the Fig. \ref{fig_airlines_data}), 
while it rather follows a power-law by adding the 
international flights 
(green line in the Fig. \ref{fig_airlines_data}).
Note that the distribution of the link lengths is obtained by 
the differential of the cumulative one and that the decay form 
of exponential or power-law is invariant.

Thus, link lengths are restricted in real systems, 
although the distribution may have some various forms 
as similar to the cases of degree distribution \cite{Amaral00}.

\begin{table}[htb]
\begin{center}
\begin{tiny}
\begin{tabular}{c|c|c|c|c} \hline
 class of SF nets. & generation rule & planar & length & models \\ \hline 
 with disadvantaged & connection of $(i,t)$ with prob. 
  & $\times$ & $\bigcirc$ & modulated \\ 
 long-range links 
  & $\Pi_{i}(t) \sim k_{i}(t) l^{\alpha}$, $\alpha < 0$ & & 
  & BA model \cite{Manna02} \cite{Brunet02} \\ 
  & connection of $(i,j)$ iff  & $\times$ & $\bigcirc$ 
  & geo. threshold \\
  & $(w_{i} + w_{j}) h(r_{ij}) \geq \theta$ & & 
  & graph \cite{Masuda05} \\ \hline
 embedded on & with randomly assigned $k_{j}$ 
  & $\times$ & $\bigtriangleup$ &  Warren et.al. \cite{Warren02}, \\
  a lattice  & restricted links & & 
  & Avraham et.al. \cite{Avraham03} \cite{Rozenfeld02} \\ 
  & in the radius $A k_{j}^{1/d}$ & & & \\ \hline
 by space-filling & triangulation & $\bigcirc$ & $\bigtriangleup$
  & Apollonian \\
  & (geo. attach. pref.) & & 
  & nets.  \cite{Zhou04}\cite{Zhou05}\cite{Doye05} \\
  & pref. attach. by & $\times$ & $\bigtriangleup$
  & growing spatial \\
  & selecting edges  & & & SF nets \cite{Manna05} \\ \hline   
\end{tabular}
\end{tiny}
\end{center}
\caption{Summary of geographical SF network models.
 The symbols $\bigcirc$, $\bigtriangleup$, and $\times$
 denote goodness levels for each property.} \label{table_models}
\end{table}

\section{Geographical SF network models}
We review geographical SF network models    
in the state-of-the-art.  
By the generation rules of networks,   
they are categorized in three classes as shown in 
Table \ref{table_models}.  
The generation rules are explained by variations in the balance of 
minimizing the number of hops between nodes 
(benefits for transfers) and the link lengths.

In this section, 
we refer to the generation rules and the power-law behavior
only in the essential forms because of the limited pages.
The properties of planarity without crossing links
and the link lengths will be discussed in the next section.

\subsection{SF networks with disadvantaged long-range links} 
The modulated Barab\'{a}si-Albert (BA) model \cite{Manna02}
and the geographical threshold graph \cite{Masuda05}
belong to the first class: 
SF networks with disadvantaged long-range links
between nodes whose positions are random on a space \footnote{To simplify
the discussion, we assume an uniformly random distribution of nodes on a
space.
However, the procedure 
can be generalized to any other distributions.}.
They are natural extensions of the previous 
non-geographical SF network models by the competition of 
preferential linking based on the degree or weight 
and the restriction of link length (distance dependence).

\subsubsection{Modulated BA model in the Euclidean space} 
\begin{figure}[htb] 
  \begin{center}
  \includegraphics[width=80mm]{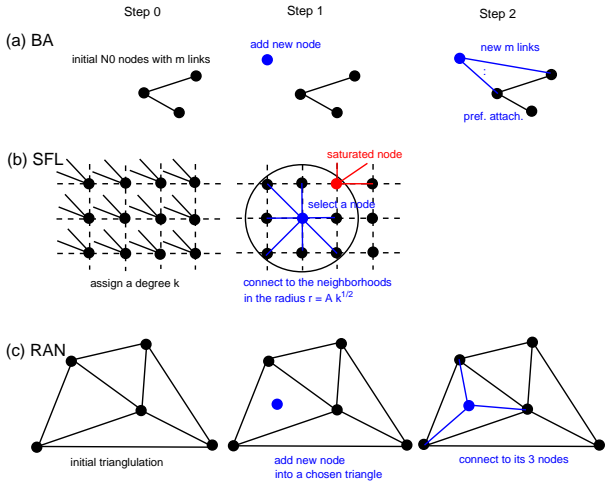}    
  \end{center}
  \caption{Network generation in each model.
 The analytically obtained 
 degree distributions for (a)-(c) follow $P(k) \sim k^{-3}$,
 $P(k) \sim k^{-\gamma}$ with cutoff $k_{c} < K$, 
 and $P(k) \sim k^{-\gamma_{RA}}$, $\gamma_{RA} \sim 3$, respectively.} 
  \label{fig_schema}
\end{figure}

Before explaining the first class, 
we introduce the well-known BA model \cite{Barabasi99}
generated by the following rule:  
{\it growth} with a new node at each time  
and {\it preferential attachment} of links to nodes with large degrees
(see Fig. \ref{fig_schema}(a)).  
\begin{description}
 \item[BA-Step 0:] A network grows from an initial $N_{0}$ nodes with    
	    $m < N_{0}$ links among them.    
 \item[BA-Step 1:] At every time step, a new node is introduced and is
	    randomly connected to $m$ previous nodes as follows.   
 \item[BA-Step 2:] Any of these $m$ links of the new node introduced at
	    time $t$ connects a previous node $i$ with an attachment   
	    probability $\Pi^{BA}_{i}(t)$ which is linearly proportional   
	    to the degree $k_{i}(t)$ of the $i$th node at time $t$,   
	    $\Pi^{BA}_{i}(t) \sim k_{i}(t)$.   
\end{description} 
The preferential attachment makes a heterogeneous network 
with hubs.
More precisely,
the degree distribution $P(k) \sim k^{-3}$ 
is analytically obtained 
by using a mean-field approximation \cite{Barabasi99}
in the continuum approach \cite{Albert02},
in which the time dependence of the degree $k_{i}$ of a given node $i$
is calculated through the continuous real variables of degree and time.

Based on a competition between the preferential attachment 
and the distance dependence of links,   
the modulated BA model on a space with physical distance has been   
considered \cite{Manna02}.  
Note that the position of new node is random.
The network is grown by introducing at unit rate randomly selected nodes  
on the Euclidean space (e.g. two-dimensional square area), and   
the probability of connection is modulated according to    
$\Pi_{i}(t) \sim k_{i}(t) l^{\alpha}$,    
where $l$ is the Euclidean distance between the $t$th    
(birth at time $t$) and the older $i$th nodes, 
and $\alpha$ is a parameter.  
The case of $\alpha = 0$ is the original BA model \cite{Barabasi99}.   
In the limit of $\alpha \rightarrow -\infty$,   
only the smallest value of $l$ corresponding to the nearest node will   
contribute with probability $1$.   
Similarly, in the limit of $\alpha \rightarrow \infty$,   
only the furthest node will contribute.   
Indeed, it has been estimated that the distribution of link lengths  
follows a power-law $l^{- \delta}$    
(long-range links are rare at $\delta > 0$),   
whose exponent is calculated by $\delta = \alpha + d - 1$  
for all values of $\alpha$ \cite{Manna02}.

In the modulated BA model on a one-dimensional lattice
(circumference),    
it has been numerically shown that \cite{Brunet02}   
for $-1 < \alpha < 0$ the degree distribution is close to a power-law,   
but for $\alpha < -1$    
it is represented by a stretched exponential 
$P(k) = a \exp( -b k^{\gamma})$, 
where the parameters $a$, $b$, 
and $\gamma$ depend on $\alpha$ and $m$, 
although the SW property \cite{Watts98}
is preserved at all values of  
$\alpha$. 
For the transition from the stretched exponential to the SF behavior, 
the critical value is generalized to 
$\alpha_{c} = 1 - d$ in the embedded $d$-dimensional space 
\cite{Manna02}. 
More systematic classification in a parameter space of the exponents of  
degree, distance, and fractal dimension has been also discussed 
\cite{Yook02}. 

Other related studies to the form of connection probability 
$\Pi_{i}\sim k_{i}^{\beta} l^{\alpha}$ 
are the phase diagram of the clustering properties in the
$\alpha$-$\beta$ plane \cite{Sen03}, the comparison of the topological
properties for the special case of the connection probability
proportional to the distance ($\alpha =1$, $\beta = 0$)
and the inverse distance ($\alpha =-1$, $\beta = 0$) \cite{Jost02}, 
the numerical investigation of the scaling for the quantities 
(degree, degree-degree correlation, clustering coefficient) of 
the network generated by the connection probability proportional to the
degree with the exponential decay of the distance \cite{Barthelemy03},
and so on.

\subsubsection{Geographical threshold graphs} 
Th geographical threshold graph \cite{Masuda05}
is a non-growing network model extended form 
the threshold SF network model \cite{Caldareli02} \cite{Masuda04}.
It is embedded in the $d$-dimensional Euclidean 
space with disadvantaged long-range links.
We briefly show the analysis of degree distribution.

Let us consider a set of nodes with the size $N$.
We assume that 
each node $i$ is randomly distributed with uniform density $\rho$
in the space whose coordinates are denoted by 
$x_{1}, x_{2}, \ldots, x_{d}$, 
and that it is assigned with a weight $w_{i} \geq 0$ 
by a density function $f(w)$.
According to the threshold mechanism \cite{Masuda05}, 
a pair of node $(i, j)$ is connected iff
\begin{equation}
  (w_{i} + w_{j}) h(r_{ij}) \geq \theta, \label{eq_cond_thresh}
\end{equation}
where $h(r_{ij})$ is a decreasing function of the distance 
$r_{ij} > 0$ 
between the nodes, and $\theta$ is a constant threshold.

If $f(w)$ is the Dirac delta function at $w* > 0$, 
then the condition of connection (\ref{eq_cond_thresh})
is equivalent to 
$r_{ij} \geq h^{-1}\left(\frac{\theta}{2 w^{*}}\right) 
\stackrel{\rm def}{=} r^{*}$
by using the inverse function $h^{-1}$.
This case is the unit disk graph,
as a model of mobile and sensor networks,
in which the two nodes within the radius $r^{*}$ are connected
according to the energy consumption.
However, 
the degree distribution $P(k)$ is homogeneous.
We need more inhomogeneous weights.

Thus, if the exponential weight distribution 
\begin{equation}
  f(w) = \lambda e^{ - \lambda w}, \label{eq_exp_weight}
\end{equation}
and the power-law decay function 
$h(r_{ij}) = (r_{ij})^{- \beta}$, $\beta \geq 0$, 
are considered, 
then the degree is derived as a function of weight 
\begin{equation}
  k(w_{i}) = \int_{0}^{\infty} f(w_{j}) d w_{j}
   \int_{(w_{i} + w_{j})/(r_{ij})^{\beta} \geq \theta} 
   \rho d x_{1} \ldots d x_{d}
   \sim e^{\lambda w_{i}} \label{eq_kw},
\end{equation}
after slightly complicated calculations.
The second integral in the r.h.s of (\ref{eq_kw}) 
is the volume of $d$-dimensional hypersphere.
As in Refs. \cite{Masuda05} \cite{Masuda04}, 
by using the relation of cumulative distributions
$\int_{0}^{k(w)} P(k) d k = \int_{-\infty}^{w} f(w') d w'$, 
we have 
\begin{equation}
  P(k) = f(w) \frac{d w}{d k}. \label{eq_pk}
\end{equation}
From (\ref{eq_kw}) and (\ref{eq_pk}), 
we obtain the power-law degree distribution 
\[
  P(k) \sim e^{- 2 \lambda w} \sim k^{-2}.
\]
Note that this result is derived 
only if the value of $\beta$ is sufficiently small, 
otherwise the degree distribution has a streched exponential decay or an
exponential decay.

On the other hand, for the power-law 
weight distribution (called Parete distribution in this form)
\begin{equation}
  f(w) = \frac{\alpha}{w^{*}} \left(
	\frac{w^{*}}{w} \right)^{\alpha+1},  \label{eq_parete_weight}
\end{equation}
we similarly obtain 
\[
  k(w) \sim w^{d / \beta}, \;\; 
  P(k) \sim k^{ -(1+\alpha \beta / d)}.
\]
The exponent $\gamma \stackrel{\rm def}{=} 1+\alpha \beta / d$
is a variable depends on the parameters 
$\alpha$ and $\beta$.

Furthermore, we mention a gravity model with 
$h(r_{ij}) = 1 / \log r_{ij}$.
In this case, 
the condition of connection (\ref{eq_cond_thresh}) is rewritten as 
$w_{i} + w_{j} \geq \theta \log r_{ij}$,
and into 
\begin{equation}
  \frac{W_{i}W_{j}}{(R_{ij})^{\beta}} \geq \theta, \label{eq_gravity}
\end{equation}
by the variable transformations 
$W_{i} \stackrel{\rm def}{=} e^{w_{i}}$, 
$W_{j} \stackrel{\rm def}{=} e^{w_{j}}$,
and $(R_{ij})^{\beta} 
     \stackrel{\rm def}{=} (r_{ij})^{\theta} / \theta$.
Eq. (\ref{eq_gravity}) 
represents a physical, sociological, or chemical interactions
with power-law distance dependence.
For the combination of (\ref{eq_gravity}) 
and the weight distributions $f(w)$ 
in (\ref{eq_exp_weight}) and (\ref{eq_parete_weight}),
we can also derive the more complicated forms of $P(k)$.
Thus, the choice of $f(w)$
matters for the SF properties 
in contrast to an approximately 
constant exponent $\gamma \approx 2$
in the non-geographical threshold graphs \cite{Masuda04}
without $h(r_{ij})$.

\subsection{SF networks embedded on lattices}
The second class is based on the 
SF networks embedded on regular Euclidean  
lattices (SFL) accounting for graphical properties   
\cite{Avraham03} \cite{Rozenfeld02}. 
We distinguish this class from the first one,
because the position of node is not randomly distributed 
but well-ordered on a lattice with a scale that 
gives the minimum distance.

Let us consider a $d$-dimensional lattice of size $R$ with periodic  
boundary conditions.   
The model is defined by the following configuration procedures 
(see Fig. \ref{fig_schema}(b)) 
on an assumption of power-law degree distribution.  
\begin{description}
 \item[SFL-Step 0:] To each node on the lattice,   
	    assign a random degree $k$ taken from the distribution   
	    $P(k) = C k^{- \lambda}$, $m \leq k \leq K$, $\lambda > 2$,    
	    (the normalization constant:   
	    $C \approx (\lambda -1) m^{\lambda -1}$ for large $K$).       
 \item[SFL-Step 1:]  Select a node $i$ at random, and connect it to its 
	    closest neighbors until its connectivity $k_{i}$ is realized, 
	    or until all nodes up to a distance,  
	    \begin{equation}
	       r(k_{i}) = A k_{i}^{1/d}, \label{eq_radius}  
	    \end{equation} 
	    have been explored:    
	    The connectivity quota $k_{j}$ of the target node $j$ is   
	    already filled in saturation. Here $A >0$ is a constant.
 \item[SFL-Step 2:] The above process is repeated for all nodes.   
\end{description}

As in Ref. \cite{Avraham03},
we derive the cutoff connectivity.
Consider the number of links $n(r)$ entering a node from a surrounding  
neighborhood of radius $r$,   
when the lattice is infinite, $R \rightarrow \infty$.  
The probability of connections between the origin and nodes at distance  
$r'$ is   
\[  
  P\left( k > \left(\frac{r'}{A}\right)^{d} \right)   
   = \int_{(r'/A)^{d}}^{\infty} P(k') dk' \sim   
   \left\{ \begin{array}{cc} 
    1                     & r' < A \\  
    (r'/A)^{d(1-\lambda)} & r' > A.  
   \end{array} \right.  
\]  
Thus, from   
$n(r) = \int_{0}^{r} S_{d} r'^{d-1} d r' 
   \int_{(r'/A)^{d}}^{\infty} P(k') dk'$, we obtain   
\[   
  n(r) = V_{d} r^{d} \left\{ \left( \frac{A}{r} \right)^{d}   
	       \int_{0}^{(r/A)^{d}} k P(k) dk   
	       + \int_{(r/A)^{d}}^{\infty} P(k) dk \right\},  
\]  
where $V_{d} = S_{d} / d$ and $S_{d}$  
is the volume and the surface area of the $d$-dimensional unit sphere,
respectively.  
The cutoff connectivity is then   
\begin{equation}
  k_{c} = \lim_{r \rightarrow \infty} n(r) =   
   V_{d} A^{d} \langle k \rangle, \label{eq_cutoff_kc}  
\end{equation} 
where $\langle k \rangle = \int k P(k) dk$    
denotes the average connectivity.   

 If $A$ is large enough such that $k_{c} > K$,    
the network can be embedded without cutoff.   
Otherwise, by substituting (\ref{eq_cutoff_kc}) into (\ref{eq_radius}),    
the cutoff connectivity $k_{c}$ implies a cutoff length    
\begin{equation}
 \xi = r(k_{c}) = (V_{d} \langle k \rangle)^{1/d} A^{2}.   
\end{equation} 
The embedded network displays the original (power-law) distribution up   
to length scale $\xi$ and repeats, statistically, at length scales  
larger than $\xi$.  

Whenever the lattice is finite, $R < \infty$,   
the number of nodes is finite,   
$N \sim R^{d}$, which imposes a maximum connectivity,   
\begin{equation}
  K \sim m N^{1/(\lambda -1)} \sim R^{d/(\lambda-1)}, \label{eq_def_K}  
\end{equation} 
where the first approximation is obtained from   
$\int_{K}^{\infty} P(k) dk = 
\left[ \frac{C}{1 - \lambda} k^{1 - \lambda} \right]_{K}^{\infty} = 1/N$.    
From (\ref{eq_radius}) and (\ref{eq_def_K}),     
a finite-size cutoff length is   
\begin{equation}
  r_{max} = r(K) \sim A R^{1/(\lambda-1)}.  
\end{equation} 

These three length scales, $R$, $\xi$, $r_{max}$,   
determine the nature of networks.  
If the lattice is finite, then the maximum connectivity   
$K$ is attained only if $r_{max} < \xi$.  
Otherwise ($r_{max} > \xi$), the cutoff $k_{c}$ is imposed.  
As long as $\min (r_{max}, \xi) \ll R$, the lattice size $R$ imposes   
no serious restrictions.  
Otherwise ($\min (r_{max}, \xi) \geq R$), finite-size effects bounded by  
$R$ becomes important.  
In this regime,   
the simulation results \cite{Avraham03} \cite{Rozenfeld02}   
have also shown that   
for larger $\lambda$ the network resembles the  
embedding lattice because of the rare long-range links,   
while the long-range links becomes noticeable as $\lambda$ decreases.  

Concurrently with the above work, Warren et al. \cite{Warren02}   
have proposed a similar embedding algorithm   
in a two-dimensional lattice, however 
the number of nodes in each circle  
is equal to the connectivity without cutoff. 
Thus, the main difference in their approaches is that   
a node can be connected to as many of its closest neighbors as  
necessary, until its target connectivity is fulfilled.   

In addition, Ref. \cite{Moukarzel02} has discussed the 
shortest paths on $d$-dimensional lattices with the addition of an
average of $p$ long-range 
bonds (shortcuts) per site, whose length $l$ is distributed
according to $P_{l} \sim l^{- \mu}$.

\begin{figure}[htb] 
  \begin{center}
  \includegraphics[width=85mm]{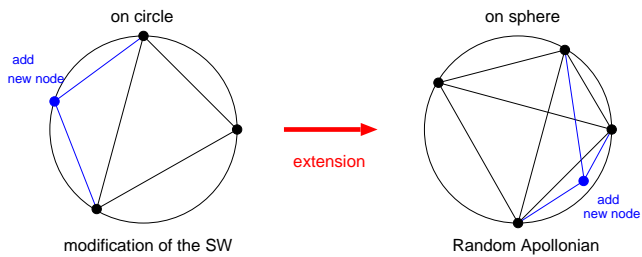}    
  \end{center}
  \caption{Growing networks with geographical attachment preference.
 The blue node and links are newly added.} 
  \label{fig_geo_attach}
\end{figure}

\subsection{Space-filling networks}
The third class is related to the space-filling packing in which 
a region is iteratively partitioned into subregions 
by adding new node and links.

\subsubsection{Growing small-world networks}
Let us consider   
the growing network with geographical attachment preference   
\cite{Ozik04} as   
a modification of the SW model \cite{Watts98}.   
In this network, from an initial configuration with $m+1$ completely  
connected nodes on the circumference of a circle, at each subsequent  
time step, a new node is added in an uniform-randomly chosen interval,  
and connects the new node to its $m$ nearest neighbors w.r.t distance  
along the circumference.  
Fig. \ref{fig_geo_attach} (left) illustrates the case of $m = 2$.  
We denote $n(k, N)$ as the number of nodes with degree $k$    
when the size (or time) is $N$.   
At time $N$, a new node with degree $m$ is added to the network, and if    
it connects to a preexisting node $i$, 
then the degree is updated by 
$k_{i} \rightarrow k_{i} + 1$ with the equal probability $m/N$      
to all nodes 
because of the uniform randomly chosen interval.   

Thus, we have the following evolution equation,      
\[  
 n(k, N+1) = \left(1 - \frac{m}{N} \right) n(k, N)      
 + \frac{m}{N} n(k-1, N) +\delta_{k,m},     
\]  
where $\delta_{k,m}$ is the Kronecker delta.    
Note that considering such equation for the average number of nodes with
$k$ links at time $N$ is called ``rate-equation approach,''
while considering the probability $p(k, t_{i}, t)$ that at time $t$
a node $i$ introduced at time $t_{i}$ has a degree $k$ 
is called ``master equation approach'' \cite{Albert02}.

When $N$ is sufficient large, $n(k, N)$ can be approximated as   
$N P(k)$. In the term of degree distribution, we obtain   
\[  
 P(k) = \frac{1}{m+1} \left( \frac{m}{m+1} \right)^{k-m},     
\]  
for $k \gg m$ ($P(k) = 0$ for $k < m$),   
although it is not a power-law.

\begin{figure}[htb]
 \begin{minipage}{.47\textwidth}
    \centering{\includegraphics[width=30mm]{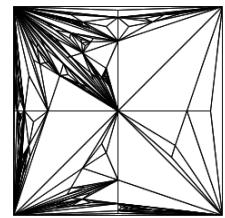}}
    \begin{center} (a) RAN \end{center}
 \end{minipage} 
 \hfill 
  \begin{minipage}{.47\textwidth}
    \centering{\includegraphics[width=35mm]{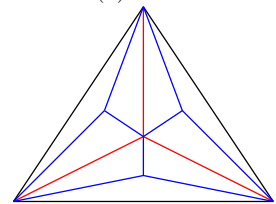}}
    \begin{center} (b) Deterministic AN \end{center}
  \end{minipage} 
  \caption{Apollonian Networks:
 (a) Random AN generated from an initial triangulation of square 
 and (b) Deterministic AN generated from an initial triangle
 of back lines.
 The red and blue lines are added links 
 at the first and second steps, respectively.} 
  \label{fig_apollonian}
\end{figure}

\subsubsection{Apollonian networks}
The growing small-world networks model \cite{Ozik04} 
can be extended from polygonal divisions on a circle   
to polyhedral divisions on a sphere   
as shown in Fig. \ref{fig_geo_attach}.  
We should remark the extended model becomes a planar graph, 
when each node on the surface is projected onto a plane such as
from a Riemannian sphere.
It is nothing but a random Apollonian network (RAN)  
\cite{Zhou04}\cite{Zhou05},  and 
also the dual version of Apollonian packing for space-filling  
disks into a sphere \cite{Doye05},   
whose hierarchical structure   
is related to the SF network formed by the minima and transition  
states on the energy landscape \cite{Doye02}.  
The power-law degree distribution 
has been analytically shown in the RAN \cite{Zhou04}\cite{Zhou05}.
To derive the distribution $P(k)$, we consider the configuration  
procedures of RAN as follows (see Fig. \ref{fig_schema}(c)). 
\begin{description}
 \item[RAN-Step 0:] Set an initial triangulation with $N_{0}$ nodes.
 \item[RAN-Step 1:] At each time step, a triangle is randomly chosen,
	    and a new node is added inside the triangle.
 \item[RAN-Step 2:] The new node is connected to its three nodes of
	    the chosen triangle. 
 \item[RAN-Step 3:] The above processes in Steps 1 and 2 
	    are repeated until reaching the required size $N$ .
\end{description}

Since the probability of connection to a node is higher as the number  
of its related triangles is larger,   
it is proportional to its degree as the preferential attachment. 
Thus, we have the following rate-equation   
\begin{equation}
 n(k+1, N+1) = \frac{k}{N_{\triangle}} n(k, N)   
 + \left( 1 - \frac{k+1}{N_{\triangle}} \right) n(k+1, N),   
 \label{eq_evol_RAN} 
\end{equation}
where the number of triangles $N_{\triangle}$     
(at the grown size or time $N$) is defined as   
$N_{\triangle} = 2(N -4) + 4$  
for an initial tetrahedron,   
and $N_{\triangle} = 2(N -3) + 1$ for an initial triangle, etc.  

In the term of $P(k) \approx n(k, N)/N$, Eq. (\ref{eq_evol_RAN})   
can be rewritten as    
\[ 
 (N+1) P(k+1) = \frac{N k P(k)}{N_{\triangle}} + N P(k+1)   
 - \frac{N (k+1) P(k+1)}{N_{\triangle}}.  
\] 
By the continuous approximation, we obtain   
the solution $P(k) \sim k^{- \gamma_{RA}}$ with   
$\gamma_{RA} = (N_{\triangle} + N)/ N \approx 3$ for large $N$.  
Fig. \ref{fig_apollonian} (a) shows an example of RAN.

Moreover, 
in the deterministic version \cite{Doye05}\cite{Andrade05}, 
analytical forms of 
the power-law degree distribution $P(k)$, 
clustering coefficient $c_{i}$, 
and the degree-degree correlation $k_{nn}(k)$ 
can be derived \cite{Doye05}, 
since the calculations are easier 
in the recursive structure without randomness into subregions 
as shown in Fig \ref{fig_apollonian} (b).
Here, 
$k_{nn}(k)$ is defined by the the average degree of the nearest 
neighbors of nodes with degree $k$.
It has been observed in technological or biological networks and in 
social networks that 
there exists two types of correlations, namely disassortative and
assortative mixings \cite{Newman03b}.
These types of networks tend to have connections between nodes 
with low-high degrees and with similar degrees, respectively.
The RAN shows the disassortative mixing \cite{Doye05}.

Similarly, the analytical forms in the high-dimensional both 
random \cite{Zhang05b} \cite{Gu05} 
and deterministic \cite{Zhang05c}
Apollonian networks have been investigated 
by using slightly different techniques.
They are more general space-filling models embedded in a 
high-dimensional Euclidean space,
although the planarity is violated.

\begin{figure}[htb]
 \begin{minipage}{.47\textwidth}
    \centering{\includegraphics[width=45mm]{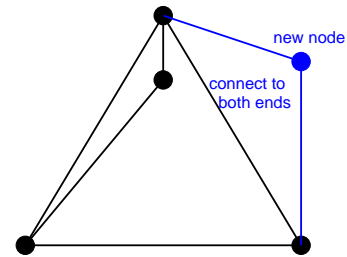}}
    \begin{center} (a) growing SW network \end{center}
 \end{minipage} 
 \hfill 
  \begin{minipage}{.47\textwidth}
    \centering{\includegraphics[width=45mm]{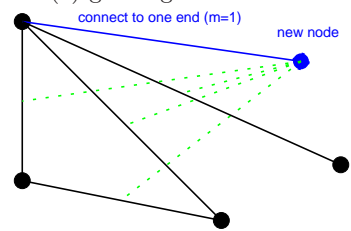}}
    \begin{center} (b) growing spatial SF network \end{center}
  \end{minipage} 
  \caption{SW and SF networks generated by randomly selecting edges.} 
  \label{fig_select_edges}
\end{figure}

\subsubsection{SF networks generated by selecting edges}
Another modification of the 
growing SW networks \cite{Ozik04} 
is based on random selection of edges \cite{Zhang05a} \cite{Manna05}.
We classify them in the relations to the partitions of interval or
region as mentioned in 3.3.1 and a Voronoi diagram.
The following two models give typical configurations
(see Fig. \ref{fig_select_edges}).

The growing SW network generated by selecting edges 
\cite{Zhang05a} is constructed as follows.
Initially, the network has three nodes, each with degree two.
As shown in Fig. \ref{fig_select_edges}(a),
at each time step, a new node is added, 
which is attached via two links
to both ends of one randomly chosen link that has never been selected
before. The growth process is repeated until reaching the required size
$N$.
Since all nodes have two candidates of links for 
the selection at each time,
an exponential degree distribution has been analytically obtained.
If the multi-selection is permitted for each link,
it follows a power-law.
The difference in configuration procedures 
for RAN is that, instead of triangulation with adding three links,
two links are added at each step.
We assume that the position of added node is random
(but the nearest to the chosen link) on a metric space, 
although only the topological properties 
are discussed in the original model \cite{Zhang05a}.

In the growing spatial SF network \cite{Manna05}
on a two-dimensional space,
$m$ links whose center points are nearest to an added  new 
node (as guided by the dashed lines in Fig. \ref{fig_select_edges} (b))
are chosen at each time step.
Both end nodes of the nearest link(s) have 
an equal probability of connection.
If a Voronoi region \cite{Imai00} \cite{Okabe00}
for the center points of links 
is randomly chosen for the position of new node in the region
\footnote{Although the position of node is randomly selected on a
two-dimensional space in the original paper \cite{Manna05}, it is
modified to the random selection of a Voronoi region which is related to
triangulation such as in RAN.
Note that it gives a heterogeneous spatial distribution of points.}, 
the selection of a link is uniformly random, 
therefore the probability of connection to each node 
is proportional to its degree.
Then, we can analyze the degree distribution.
Note that any point in the Voronoi region is closer to 
the center (called generator point) belong in it 
than to any other centers.

For the case of $m = 1$ as a tree,
the number of node with degree $k$ is evolved in the rate-equation 
\begin{equation}
  n(k, t+1) = n(k, t) + \frac{(k-1)}{2 t} n(k-1, t) 
  - \frac{k}{2 t} n(k, t) + \delta_{k,1}, \label{eq_spatial_SF}
\end{equation}
where $n(k, t)$ denotes the number of nodes with degree $k$,
and $2 t$ is the total degree at time $t$.

In the term of degree distribution 
$P(k, t) \approx n(k, t)  / t$ at time $t$, 
Eq. (\ref{eq_spatial_SF}) is rewritten as 
\[
  (t + 1) P(k, t+1) - t P(k, t) = 
\]
\[
  \frac{1}{2} 
  \left[ (k-1) P(k-1, t) -k P(k, t)
  \right] + \delta_{k,1}.
\]
At the stationary value independent of time $t$, we have 
\[
 P(k) = \frac{1}{2} 
  \left[ (k-1) P(k-1) -k P(k)
  \right] + \delta_{k,1}.
\]
From the recursion formula 
and $P(1) = 2/3$, we obtain the solution
\[
  P(k) = \frac{k-1}{k+2} P(k-1) 
  = \frac{4}{k(k+1)(k+2)} \sim k^{-3}.
\]

\section{Relations among the models}
We discuss the relations among the independently proposed models.
Remember the summary of the geographical SF network models 
in Table \ref{table_models}.

The first class is based on a combination of the preferential attachment 
or the threshold mechanism and the penalty of long-range links between
nodes whose position is random,
while the second one is on embedding the SF structure with a given 
power-law degree distribution in a lattice.
Since the assigned degree to each node 
can be regarded as a fitness \cite{Albert02}, 
the SFL is considered as 
a special case of the fitness model \cite{Caldareli02}
embedded on a lattice.
In contrast, 
the penalty of age or distance dependence of each node 
can be regarded as a non-fitness in general term.
If we neglect the difference of penalties,
this explanation bridges 
the modulated BA \cite{Manna02}\cite{Brunet02}, 
SFL \cite{Avraham03}\cite{Rozenfeld02}, 
and aging models \cite{Dorogovtsev00} 
with a generalized fitness model.
The crucial difference is the positioning of nodes:
one is randomly distributed on a space and another is well-ordered on a
lattice with the minimum distance between nodes.
Moreover, 
the weight in the threshold graphs \cite{Masuda05}\cite{Masuda04}
is corresponded to a something of fitness,
however the deterministic threshold 
and the attachment mechanisms should be
distinguished in the non-growing and growing networks.
We also remark that, in the third class, 
the preferential attachment is implicitly performed,
although the configuration procedures are more geometric based on 
triangulation \cite{Zhou04}\cite{Zhou05}\cite{Doye05}
or selecting edges \cite{Manna05}.
In particular, 
the position of nodes in the Apollonian networks 
is given by the iterative subdivisions
(as neither random nor fixed on lattice), 
which may be related to 
territories for communication or 
supply management in practice.

Next, we qualitatively compare the properties of planarity without
crossing links and link lengths.
We emphasize that the planarity is important and natural requirement
to avoid 
the interference of beam (or collision of particles)
 in wireless networks, airlines, 
layout of VLSI circuits, vas networks clinging to cutis, 
and other networks on the earth's surface, etc \cite{Zhou05}.

In the modulated BA models and the geographical threshold graphs, 
long-range links are restricted by the strong constraints
with decay terms,
however crossing links may be generated.
There exist longer links from hubs in the SFL,
because such nodes have large number of links 
however the positions of nodes are restricted on a lattice;
the density of nodes is constant, therefore they must connect to some 
nodes at long distances.
More precisely, it depends on the exponent $\lambda$ of 
power-law degree distribution as mentioned in the subsection 3.2.
In addition, 
the planarity is not satisfied by the crossing between the 
lattice edges and the short-cuts.
On the other hand, 
RAN has both good properties of the planarity and averagely short
links.
However, in a narrow triangle region, 
long-range links are partially generated as shown in 
Fig. \ref{fig_apollonian}.
Similarly, 
the SF networks generated by selecting edges may have 
long-range links as shown in Fig. \ref{fig_select_edges} (b): 
the chosen end point for connection is 
long away from the newly added node at a random position, 
even though the selected edges have the nearest centers.

\section{Conclusion}
In this review of geographical SF network models, 
we have categorized them in three
classes by the generation rules: disadvantaged long-range links, 
embedding on a lattice, and space-filling.
We have shown that these models have 
essential mechanisms to generate power-law degree distributions,
whose analytical forms can be derived on an assumption of the restricted 
link lengths as consistent with real data.
Furthermore, 
the basic topological properties of the planarity and link length 
have been discussed for each model.
In particular, the geographical threshold graphs 
and the RAN are attractive because of 
the tunable exponent $\gamma$ of $P(k) \sim k^{-\gamma}$
or the locality related to the unit disk graphs, and 
the planarity of network on the heterogeneous positioning of nodes.
However, they have drawbacks of crossing and long-range links,
respectively.
To avoid long-range links,
an improvement by the combination of RAN and Delaunay triangulation 
based on diagonal flipping \cite{Imai00}\cite{Okabe00} 
is considering \cite{Hayashi05}.

We have grasped 
several configuration procedures of geographical SF networks 
and discussed the above properties, however 
these are still at the fundamental level.
We must consider further issues, for example, 
\begin{itemize}
  \item Quantitative investigation of the topological properties 
	including diameter of network, clustering coefficient,
	degree-degree correlation, and betweenness centrality
	(related to congestion of information flow), etc. 
  \item Analysis of dynamics for the traffic and the fault-tolerance, 
	especially in disaster or emergent environment.
  \item Positioning of nodes with aggregations according to a population
	density in the evolutional and distributed manners.
\end{itemize}

We will progress to the next stage from the observation of real networks
to the development of future networks.
The distributed design and management will be usefully applied to 
many socio-technological infrastructures.

\bibliographystyle{ipsjsort}

\end{document}